\documentclass[12pt]{article}

\usepackage{amsmath,amsfonts}
\usepackage{amssymb}
\usepackage{epsfig,cite}
\usepackage{stmaryrd}
\usepackage[usenames,dvips]{color}
\usepackage{comment}
\usepackage{psfrag}
\usepackage{fullpage}

\makeatletter
\@addtoreset{equation}{section}
\makeatother

\definecolor{pink}{rgb}{.043,.63,.63}

\newcommand{\be}{\begin{equation}}
\newcommand{\ee}{\end{equation}}
\newcommand{\bea}{\begin{eqnarray}}
\newcommand{\eea}{\end{eqnarray}}
\newcommand{\Tr}{\operatorname{Tr}}
\newcommand{\tr}{\operatorname{tr}}
\newcommand{\diag}{\operatorname{diag}}

\newcommand{\nn}{\nonumber}

\begin{document}

\thispagestyle{empty}

\begin{center}
\hfill BI-TP 2012/04\\
\hfill UAB-FT-700
\begin{center}

\vspace{.5cm}

{\LARGE\sc  The Light Stop Scenario from \\ \vspace{.4cm}Gauge Mediation }

\end{center}

\vspace{1.cm}

\textbf{ Antonio Delgado$^{\,a}$, Germano Nardini$^{\,b}$
and Mariano Quir\'os$^{\,a,c}$}\\

\vspace{1.cm}
${}^a\!\!$ {\em {Department of Physics, University of Notre Dame\\Notre Dame, IN 46556, USA}}

\vspace{.1cm}
${}^b\!\!$ {\em {Fakult\"at f\"ur Physik, Universit\"at Bielefeld,
    D-33615 Bielefeld, Germany}}

\vspace{.1cm}

${}^c\!\!$ {\em {Instituci\'o Catalana de Recerca i Estudis  
Avan\c{c}ats (ICREA) and\\ Institut de F\'isica d'Altes Energies, Universitat Aut{\`o}noma de Barcelona\\
08193 Bellaterra, Barcelona, Spain}}

\end{center}

\vspace{0.8cm}

\centerline{\bf Abstract}
\vspace{2 mm}
\begin{quote}\small
  In this paper we embed the light stop scenario, a MSSM framework
  which explains the baryon asymmetry of the universe through a strong
  first order electroweak phase transition, in a top-down
  approach. The required low energy spectrum consists in the light
  SM-like Higgs, the right-handed stop, the gauginos and the Higgsinos
  while the remaining scalars are heavy. This spectrum is naturally
  driven by renormalization group evolution starting from a heavy
  scalar spectrum at high energies.  The latter is obtained through a
  supersymmetry-breaking mix of gauge mediation, which provides the
  scalars masses by new gauge interactions, and gravity mediation,
  which generates gaugino and Higgsino masses. This supersymmetry
  breaking can also explain the $\mu$ and $B_\mu$ parameters necessary
  for electroweak breaking and predicts small tri-linear mixing terms
  $A_t$ in agreement with electroweak baryogenesis requirements. The
  minimal ultraviolet embedding predicts a Higgs mass around its experimental
  lower bound and by a small extension higher masses
  $m_H\lesssim 127$ GeV can be accommodated. \\
\end{quote}

\vfill

 \newpage

\section{\sc Introduction}
\label{introduction} 

Understanding the baryon-to-entropy ratio is one of the big challenges
of contemporary Particle Physics. A particularly appealing solution to
the problem, dubbed baryogenesis, was proposed many years ago by
Sakharov~\cite{Sakharov:1967dj} who formulated the necessary
conditions a theory should exhibit to generate the baryon asymmetry: i)
$B$ violation; ii) $C$ and $CP$ violation; and, iii) Departure from thermal
equilibrium.  Later, in 1985, Kuzmin, Rubakov and
Shaposhnikov~\cite{Kuzmin:1985mm} realized that all three Sakharov's
conditions could be satisfied during a {\it strong} first-order
electroweak phase transition, and in 1993 Cohen, Kaplan and
Nelson~\cite{Cohen:1993nk} calculated the
baryon asymmetry that is produced by the $CP$ violating interactions of
fermions with the bubble domain walls.  This baryogenesis scenario
was called electroweak baryogenesis (EWBG) and in fact this mechanism
involves only physics at the electroweak scale that colliders like the LHC
can probe.

Although the Standard Model (SM) fulfills all Sakharov's conditions it
fails quantitatively~\cite{AndH,improvement,twoloop,nonpert,CPSM}
because: i) The $CP$ violation provided by the CKM phase is too small
to generate the required baryon asymmetry; ii) The phase transition is
not strong enough since it is triggered by the electroweak gauge
bosons. As a result EWBG requires extensions of the SM with scalars
strongly coupled to the Higgs sector. One of the best motivated of
these theories is the minimal supersymmetric extension of the Standard
Model (MSSM) where stops are coupled to the Higgs sector with
top-strength. However the MSSM with generic values of the supersymmetry
breaking parameters in agreement with present experimental bounds also
fails to reproduce the observed baryon asymmetry~\cite{early} since it
essentially reproduces the SM results.

A window in the space of the MSSM supersymmetry breaking parameters
where the generated baryon asymmetry can reproduce the observed values
was found in
Refs.~\cite{CQW,Delepine,CK,FL,JoseR,JRB,Carena:1997gx,Carena:1997ki,CJK,
  Iiro2,Toni2,Worah,Schmidt,Cline:2000kb,
  Carena:2000id,Konstandin:2005cd,Cirigliano:2006dg} and the model was
subsequently dubbed Light Stop Scenario
(LSS)~\footnote{\label{footnoteLSS}There are two main qualitatively
  and quantitatively different ways of achieving the LSS: i) By
  considering at low energies $m_U\simeq m_Q\sim{\mathcal O}($TeV) and
  $|A_t/m_Q|^2\gg 1$; and, ii) By considering at low energies $m_Q$ in
  the multi-TeV region, to cope with present experimental bounds on
  the Higgs mass, and $m_Q^2\gg m_U^2$ and $|A_t/m_Q|^2\ll 1$. In this
  paper we will call LSS the second scenario where EWBG in agreement
  with observations can be enforced.}. The main lines of the LSS go as
follows. If the left-handed stop doublet $Q$ is heavy enough not to
contribute too much to electroweak observables (particularly the
$\rho$ parameter) and to sufficiently increase the Higgs mass by
radiative corrections, then the right-handed stop singlet $U$ can be
light and trigger a strong first-order phase transition provided that:
i) The supersymmetry breaking mass parameter $m_U^2$ and the thermal
mass at the critical temperature $T_c\simeq 100$ GeV (nearly) cancel
to each other~\footnote{Of course a tachyonic mass for the
  right-handed stop creates a charge and color breaking minimum. The
  tree-level condition for the electroweak minimum to be the true one
  at zero temperature is~\cite{CQW} $\left| m_U^2\right|\lesssim m_H v
  g_3/\sqrt{12}\simeq (100\, \textrm{GeV})^2$.  This condition is
  modified when a metastable but long-lived electroweak minumum is
  allowed and radiative and/or thermal corrections are
  added~\cite{Carena:2008rt}. }~\cite{CQW}, i.e. $m_U^2\simeq -
m_Z^2$; and, ii) The mixing parameter $A_t$, which penalizes the
strength of the phase transition ~\cite{CQW}, is small compared to the
left-handed stop mass, i.e. $A_t \lesssim
m_Q/2$~\cite{Carena:2008rt}. Consequently the physical right-handed
stop mass has to be lower than the top-quark one.

More precisely due to the smallness of the mixing parameter and
right-handed stop mass, the left-handed stop mass must be larger than
6 TeV to cope with present limits on the Higgs mass~\cite{Carena:2008rt}.
A large value of $\tan\beta$ helps with the Higgs mass although its
value is bounded to $\tan\beta\lesssim 15$ by electric dipole moment
(EDM) observables and sufficient baryon asymmetry
production~\cite{Carena:2008rt}. The rest of third-generation
sfermions do not play any role in the phase transition and they are
usually assumed to be as massive as the first and second generation
sfermions that are kept heavy to fulfill bounds from flavour-changing
neutral currents and $CP$ violating operators. Finally a heavy
non-SM-like Higgs sector also favors the phase transition so that one
can assume all scalars, except for the right-handed stop and the SM-like Higgs, to have
masses of order (or larger than) $m_Q$. Concerning the fermionic sector, electroweak
gauginos and Higgsinos are light (say lighter than 300 GeV) to be in
thermal equilibrium with the thermal plasma and produce the required
baryon asymmetry in the bubble walls, and gluinos are at the TeV scale
not to generate a too large thermal mass for right-handed stops. This
fermionic hierarchy is fully consistent with the stronger
renormalization of gluinos from QCD interactions. In summary the effective LSS
theory at the energy scale $\mathcal{Q}\ll m_Q$ consists in the SM
fermions and Higgs, gauginos and Higgsinos and the right-handed stop.

While plenty of studies of EWBG have been performed in the
LSS~\cite{CQW,Delepine,CK,FL,JoseR,JRB,Carena:1997gx,Carena:1997ki,CJK,
  Iiro2,Toni2,Worah,Schmidt,Cline:2000kb,
  Carena:2000id,Konstandin:2005cd,Cirigliano:2006dg} a supersymmetry
breaking mechanism leading to the required soft-breaking parameters is
lacking in the literature. It is the aim of the present paper to fill
this gap and propose a plausible high-energy scenario of supersymmetry
breaking giving rise to the LSS.

The content of this paper is as follows. In Sec.~\ref{bottom} we will
determine the conditions of electroweak symmetry breaking consistent
with the LSS. These conditions will impose some relations on the
soft-breaking terms that at the scale $\mathcal{Q}\sim 10^{15}\,$GeV
manifest a peculiar hierarchy. This particular pattern will hint to
the supersymmetry breaking framework considered in
Sec.~\ref{LSSmodel}. In particular, depending on the value of the
SM-like Higgs mass $m_h$, we will propose two MSSM ultraviolet (UV)
completions consistent with the LSS: a minimal extension with
$m_h\lesssim 115.5 \,\rm GeV$ (Sec.~\ref{minimal}) and a non-minimal
one compatible with $m_h\lesssim 130 \,\rm GeV$
(Sec.~\ref{nominimal}). Finally Sec.~\ref{conclusions} is devoted to
summarize the main results and highlight future research prospects.

\section{\sc The Light Stop Scenario}
\label{bottom}
As we have described in the previous section at the scale
$\mathcal{Q}\simeq m_Z$ the LSS consists in the SM fields with a light
Higgs ($h$), light right-handed stop ($U$), gauginos and Higgsinos
eigenstates ($\chi^\pm_i,\ \chi^0_i$) with masses (including the $\mu$
parameter) of the order of the electroweak scale and the gluino ($\tilde g$) with TeV mass. The left-handed stop
($Q$), non-SM-like scalar ($H$) and pseudoscalar ($A$) Higgses as well as
right-handed sbottom, third generation sleptons and first and second
generation sfermions do not appear at low energy and we can assume
their masses in the multi-TeV range. Moreover we will work in the large
$\tan\beta$ regime, i.e. $\tan\beta\gtrsim10$, preferred by EWBG and
Higgs mass constraints \cite{Carena:2008rt}.

In this section we will study two issues of the LSS: i) How the
electroweak breaking proceeds, which will provide the boundary
conditions for the RGE, and; ii) The renormalization group evolution
of the soft-breaking terms, which will yield the spectrum at
high-scales and will provide a hint of the fundamental theory of
supersymmetry breaking at the origin of such low energy scenario.

\subsection{\sc Electroweak breaking}
\label{electroweak}
We will use the notation $H_u$ ($H_d$) for the MSSM Higgs that couples
to the top (bottom) quark via supersymmetric Yukawa coupling $h_t$
($h_b$) and acquires the vacuum expectation value $v_u$ ($v_d$)
satisfying the relations $v=\sqrt{v_u^2+v_d^2}=174$ GeV and
$\tan\beta\equiv v_u/v_d$. The necessary condition for electroweak
breaking is
\be
B_\mu^2>(m_{H_u}^2+|\mu|^2)(m_{H_d}^2+|\mu|^2) ~,
\label{ewsb}
\ee
and the minimization conditions on the Higgs potential can be written, for $m_{H_d}^2>m_{H_u}^2$, as
\begin{eqnarray}
m_{H_u}^2&=&\frac{m_{H_d}^2}{\tan^2\beta}-\frac{\tan^2\beta-1}{\tan^2\beta}\left(|\mu|^2+\frac{m_Z^2}{2}\right)\simeq \frac{m_{H_d}^2}{\tan^2\beta}-|\mu|^2-\frac{m_Z^2}{2}~,
\label{mhu}\\
B_\mu&=&\frac{1}{\tan\beta}\left(m_{H_d}^2+|\mu|^2-\frac{m_Z^2}{2}\frac{\tan^2\beta-1}{\tan^2\beta+1}\right)
\simeq
\frac{1}{\tan\beta}\left(m_{H_d}^2+|\mu|^2-\frac{m_Z^2}{2}\right) ~,
\label{Bmu}
\end{eqnarray}
where the last expressions correspond to the limit where $\tan^2\beta\gg 1$.  

From Eq.~(\ref{mhu}) we see that in the usual case with
$m_{H_d}^2\simeq |\mu|^2$ and $\tan\beta\gg 1$ the condition
(\ref{ewsb}) is automatically satisfied because the term
$(m_{H_d}^2/\tan^2\beta)$ is negligible and thus $m_{H_u}^2\simeq
-|\mu|^2-m_Z^2/2<0$. Instead, in the most promising parameter region of
the LSS it turns out that $\tan\beta\gg1$ and $m_{H_d}^2/\tan^2\beta\gg
|\mu|^2$ implying that electroweak breaking happens with
$m_{H_u}^2\simeq m_{H_d}^2/\tan^2\beta>0$ and $B_\mu^2\simeq
m_{H_d}^4/\tan^2\beta>m_{H_u}^2m_{H_d}^2$.  In summary the electroweak
symmetry breaking for the LSS happens with the hierarchy of parameters
\be
m_Z^2,\, |\mu|^2\ll m_{H_u}^2\simeq \frac{m_{H_d}^2}{\tan^2\beta}\ll
B_\mu\simeq \frac{m_{H_d}^2}{\tan\beta}\ll m_{H_d}^2 ~.
\label{hierarchy}
\ee
In fact the pseudoscalar ($A$), heavy scalar ($H$) and charged scalar
($H^\pm$) tree-level masses are degenerate and heavy:
\be
m_A^2\simeq m_{H^0}^2\simeq m_{H^\pm}^2\simeq m_{H_d}^2 ~.
\label{scalars} 
\ee

The minimum condition for the Higgs sector (\ref{mhu}), as well as the
constraint on the right-handed stop
\be
m_U^2\simeq -m_Z^2~,
\label{mU}
\ee
will be used as boundary conditions at the weak scale
$\mathcal{Q}\simeq m_Z$ for the renormalization group evolution of the
squared scalar masses towards high scales. As we will prove, such high
energy spectrum can be provided by the theory of supersymmetry
breaking.

\subsection{\sc Renormalization group running}
\label{renormalization}
Using the boundary conditions (\ref{mhu}) and (\ref{mU}) one can run
the mass spectrum from the electroweak ($\mathcal{Q}\sim m_Z$) to high
($\mathcal{Q}\gtrsim 10^{15}$ GeV) scales and try to figure out the theory
providing the corresponding soft supersymmetry breaking. In the LSS
the relevant terms in the RGE $\beta$-functions are given
by~\cite{Martin:1993zk}
\begin{eqnarray}
\label{RGE}
16\pi^2\beta_{m_Q^2}&=&2 h_t^2 X_U+2 h_b^2 X_D~,\nn\\
16\pi^2\beta_{m_U^2}&=&4 h_t^2 X_U~,\quad 16\pi^2\beta_{m_D^2}=4 h_b^2 X_D~,\\
16\pi^2\beta_{m_{H_u}^2}&=&6 h_t^2 X_U~,\quad 16\pi^2\beta_{m_{H_d}^2}=6 h_b^2 X_D~,\nn
\end{eqnarray}
where all squark masses refer to the third generation and, although
not explicitly written, $\beta_{h_t},\beta_{h_b}$ and $\beta_{g_3}$
are taken into account. In the RGE (\ref{RGE}) we have used the
definitions
\be
\label{Xs}
X_U= m_Q^2+m_{H_u}^2+m_U^2~,\quad X_D= m_Q^2+m_{H_d}^2+m_D^2~,
\ee
and $\beta_X=dX/dt$ with $t=\log (\mathcal{Q}/m_Z)$. Moreover we have
neglected the squark-mass mixings $A_{t,b}$~\footnote{As already stated EWBG
  in the MSSM requires $A_t\lesssim m_Q/2$ \cite{Carena:2008rt}.},
as well as gaugino (in particular gluino) masses which are at the electroweak scale as we are assuming that scalar masses are in the multi-TeV range. Their effect will be evaluated later on and indeed proved to be negligible.
We have kept the contribution from the bottom Yukawa coupling $h_b$,
which can be relevant only for very large values of $\tan\beta$,
although for the considered values its influence in the RGE is tiny.

Notice that once we assume that at the high scale $m_Q^2\simeq m_U^2$,
with $m_Q$ in the multi-TeV region, we can neglect from the RGE the
contribution from gaugino masses as in Eq.~(\ref{RGE}) and the LSS
condition $m_U^2 \ll m_Q^2$ is driven naturally at low energies by the
different coefficients of the squared top Yukawa coupling.

Eqs.~(\ref{RGE}) have the three independent RGE invariants:
\begin{eqnarray}
I_Q(t)&=& 3 m_Q^2-m_{H_u}^2-m_{H_d}^2\nn~,\\
I_U(t)&=& 3 m_U^2-2 m_{H_u}^2~,\\
I_D(t)&=& 3 m_D^2-2 m_{H_d}^2~,\nn
\label{I}
\end{eqnarray}
whose values are fixed by the boundary conditions of the involved
masses. They can be used to extract some analytical results as we will
see now. In the LSS it is assumed $m_Q^2(0)\gg m_U^2(0)$. Moreover $m_U^2(t)$
runs faster (with steeper slope) than $m_Q^2(t)$ because
$\beta_{m_U^2}>\beta_{m_Q^2}$. Consequently there exists a point
$t=t^*\gg1$ such that
\be
m^2_Q(t^*)=m_U^2(t^*)~,
\label{QU}
\ee
and since $t^*\gg 1$ and $m_Q^2(0)\gg m_U^2(0)$ it turns out that
$m_U^2(t^*), m_{H_u}^2(t^*)\gg m_{H_u}^2(0)$. Hence the boundary
conditions (\ref{mhu}) and (\ref{mU}) and the invariant $I_U$ imply
\footnote{In the approximations of this section terms of $\mathcal O
  (m_{H_u}^2\!(0))$ are neglected since $m_{H_u}^2\!(0)\ll m_{Q}^2\!(0) $.}
\be
m_U^2(t^*)\simeq \mathcal{R}_m\ m_{H_u}^2(t^*)~ ,
\label{UHu}
\ee
with $\mathcal{R}_{m}=2/3$. We can now make the plausible assumption that there exists a
supersymmetry breaking mechanism that provides the
equality (\ref{UHu}) while giving the same mass to both Higgs fields
$H_u$ and $H_d$. Such scenario imposes, on top of the previous
boundary conditions (\ref{mhu}), (\ref{mU}) and (\ref{QU}), the
further condition
\begin{eqnarray}
  m_{H_d}^2(t^*)=m_{H_u}^2(t^*)~.
\label{bc}
\end{eqnarray}
Notice that for any gauge mediated supersymmetry breaking mechanism based on the gauge group $G$ (see next section) condition (\ref{bc}) is implied by the existence of the $\mu H_u H_d$ term in the superpotential. In fact if $H_uH_d$ is invariant under $G$, $H_u$ transforms as the complex conjugate representation of $H_d$, and both receive the same mass at the supersymmetry breaking scale.

By plugging Eqs.~(\ref{UHu}) and (\ref{bc}) in the invariant $I_Q$ one
obtains $I_Q(0)=I_Q(t^*)\simeq 0$ and therefore the boundary
conditions at $t=0$
\be
m_{H_d}^2(0)\simeq 3m_Q^2(0)
\label{cero}
\ee
and at $t=t^*$
\be
3m_Q^2(t^*)\simeq 2m_{H_d}^2(t^*)\simeq 2m_{H_d}^2(0)
\label{tstar}
\ee
follow. The last relation in Eq.~(\ref{tstar}) is due to the fact that
$m_{H_d}^2(t)$ runs very little (just from the coupling
$h_b$). One can then use Eqs.~(\ref{cero}) and (\ref{tstar}) to
evaluate the running of the parameter $m_Q^2(t)$ between $t=0$ and
$t=t^*$ as
\be
m_Q^2(t^*)\simeq 2 m_Q^2(0)~.
\label{running}
\ee 
In fact using Eq.~(\ref{running}) one can relate the
masses provided by supersymmetry breaking at the scale $t=t^*$,
e.g. the masses in Eq.~(\ref{bc}), with the physical quantity
$m_Q(0)$, the mass of the left-handed stop doublet in the LSS.
Finally the squared mass parameter $m_D^2(t)$ runs also very slowly and
it is subject to the invariant $I_D$ in Eq.~(\ref{I}). Its value at
$t=t^*$ shall be provided by the theory of supersymmetry breaking as
we will see in the next section.

\begin{figure}[tb]
\begin{center}
\vspace{.2cm}
\includegraphics[width=0.7\textwidth]{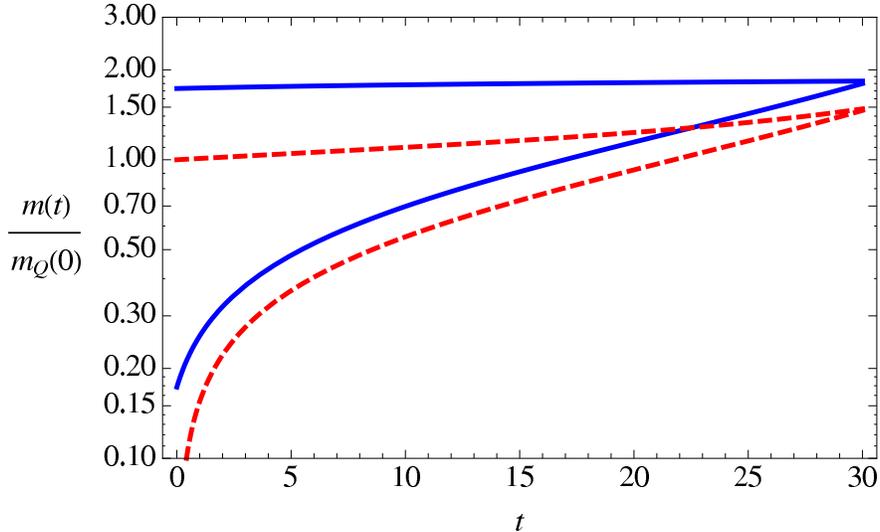}
\end{center}
\caption{\it Running masses, in units of $m_Q(0)$, as a function of
  $t=\log (\mathcal{Q}/m_Z)$ for $\tan\beta=10$. The lines are for
  $H_u$ (lower solid), $H_d$ (upper solid), $U^c$ (lower dashed) and
  $Q$ (upper dashed).  }
\label{plot}
\end{figure}

We have solved numerically the RGE (\ref{RGE}) for the supersymmetry
breaking mass parameters fulfilling the boundary conditions
(\ref{mhu}), (\ref{mU}), (\ref{QU}) and (\ref{bc}). The outcome is
presented in Fig.~\ref{plot}.
We can see from it that the scale where both $m_{H_u}^2$ and
$m_{H_d}^2$, and $m_Q^2$ and $m_U^2$ unify is $t^*\simeq 30.2$ so that
$M_*= m_Z e^{t^*}\simeq 1.2\times 10^{15}$ GeV represents the scale at
which supersymmetry breaking takes place\footnote{For
  the extreme case $A_t\approx m_Q/2$ the RGE evolution of the
  soft-mass terms is faster and $t^*$ (defined later in this section)
  would be $t^*\simeq 23$. However here we focus on $A_t/m_Q\approx 0$
  since it is the typical case emerging from the
  supersymmetry-breaking models we will consider in the next section.}. This result strongly
suggests a mechanism where supersymmetry breaking is mediated by gauge
interactions through some messengers with masses $M_*$, as it happens
in the framework of the next section.

Inclusion of TeV gluinos in the RGE of Eq.~(\ref{RGE}) amounts to the addition of the $\beta$-function terms~\cite{Martin:1993zk}
\be
16\pi^2\Delta\beta_{m^2_{X}}=-\frac{32}{3}g_3^2M_3^2,\quad X=Q,U,D
\label{gluino}
\ee
Assuming a TeV Majorana mass gluino at low energy $M_3(0)$ the term (\ref{gluino}) leads to a negligible shift of the scale $t^*$ ($\Delta t^*$) as previously anticipated. In particular for $m_Q\gtrsim 10\ (100)$ TeV we obtain $\Delta t^*\lesssim 0.1\ (0.01)$.

\section{\sc Top-down approach for the LSS}
\label{LSSmodel}
In this section we will prove the existence of a fundamental
high-energy model whose supersymmetry breaking produces a low energy
effective theory with the LSS features. We will find two possible
ultraviolet completions fulfilling all requirements: a minimal one
where successful EWBG implies $m_h\lesssim 115.5$\,GeV, border line
with the recent exclusion limits by LHC, and a non-minimal version
compatible with the whole LHC experimental band $115.5$ GeV $\lesssim
m_h\lesssim 127\,$GeV \cite{lhc}. In both models supersymmetry
breaking is gauge mediated~\cite{gmsb} at the scale $M_*$ from a
hidden sector to the observable sector by messengers
$\Phi_i+\bar\Phi_i$ with supersymmetric mass $M_*$. As it is usually done 
soft terms are introduced by a spurion field $X=\theta^2 F$, with a
supersymmetry breaking scale  $F$ such that $\Lambda\equiv F/M_*\ll M_*$, and its
transmission to the observable sector occurring by gauge interactions
and mediated by the messenger sector. Of course given the
supersymmetry breaking scale $F$ the vanishing of the cosmological
constant will imply a massive gravitino $m_{3/2}$ and gravitational
interactions will also transmit supersymmetry breaking to the
observable sector as we will discuss later on.

\subsection{\sc A Minimal UV completion}
\label{minimal}
In this section we will present a minimal theory of supersymmetry
breaking at the UV (high) scale $M_*$ which provides at the
low scale $\mathcal Q\simeq m_Z$ the main features depicted in
previous sections.
\subsubsection{\sc The scalar sector: gauge mediation}
\label{scalar}
As the scalar sector of the LSS is much heavier than the Higgsino and
gaugino sector it is clear that the gauge group responsible through
gauge mediation of the squark and slepton masses should \textit{not}
be a subgroup of the SM gauge group. We will then restrict ourselves
for simplicity to gauge supersymmetry-breaking mediators charged under
extra $U(1)$ factors.

As we want the charge $Q$ of each extra $U(1)$ to be orthogonal with
respect to the SM hypercharge (i.e. $\Tr Q\cdot Y=0$) and anomaly free
with the MSSM matter content (possibly including the right-handed
neutrino), we are led to the unique group
$U(1)_\chi$~\cite{Langacker:2008yv} that appears for instance in the
breaking $SO(10)\to SU(5)\otimes U(1)_\chi$~\footnote{Other
  anomaly-free $U(1)'s$ with the MSSM matter content as
  $U(1)_{3R}\subset SU(2)_R$ and $U(1)_{B-L}$ are not orthogonal to
  the SM-hypercharge.}, as well as in higher rank group breakings as
$E_6$~\cite{paco}.  This breaking contains vector-like extra matter so
that, since we are not concerned by unification of this group with the
SM, we can just consider the MSSM particle content plus the
right-handed neutrino field $N^c$. Moreover Eqs.~(\ref{QU}) and
(\ref{bc}), which require $Q$ and $U^c$, and $H_u$ and $H_d$ to be
equally charged (in absolute value), are satisfied by $U(1)_\chi$, as
we can see in the first row of Tab.~1 where the $Q_\chi$ charges of
these superfields are listed. The global normalization of the charges
$N_\chi$ is arbitrary since any rescaling of it can be absorbed into a
redefinition of the corresponding gauge coupling constant
$g_\chi$~\footnote{For instance the normalization which appears from the $SO(10)$ breaking is $N_\chi=2\sqrt{10}$. We will keep it arbitrary since physical results should not dependent on it.}.
\begin{table}[hbt]
\label{tabla}
\begin{center}
\begin{tabular}{||c||c|c|c|c|c|c|c|c||}\hline
Field&$Q$&$U^c$&$D^c$&$L$&$E^c$&$N^c$&$H_u$&$H_d$\\
\hline
$N_\chi Q_\chi$&-1&-1&3&3&-1&-5&2&-2\\
\hline
$N_F Q_F$&$Y_F$&$-Y_F$&$-Y_F$&$Y_F$&$-Y_F$&$-Y_F$&0&0\\
\hline
\end{tabular}
\caption{\it $U(1)_\chi$ and $U(1)_F$ charges of MSSM fields}
\end{center}
\end{table} 

Of course the $U(1)_\chi$ group is not enough to describe the spectrum
provided by Fig.~\ref{plot} and Eq.~(\ref{UHu}) because the ratio of
the Higgs over the squark squared masses should be 3/2 instead of 4 as
implied solely from $U(1)_\chi$. This means one needs to introduce an
extra $U(1)_F$ gauge group. In the absence of anomalies, and with the
field content of Table 1, $U(1)_F$ can only be a family group where
the anomaly cancelation is provided by the different generations of
fermions. If we consider fermions as triplets in the $SU(3)$ symmetry
between families, the only $U(1)$ subgroup which do not split the
masses of the first two generations while providing a mass to the
third generation is spanned by the $SU(3)$ generator
$T_8$. Consequently $Y_F$ in Tab.~1 is
\be
Y_F=\diag\left(\frac{1}{2},\frac{1}{2},-1\right)~,
\label{YF}
\ee
where the i-th entry corresponds to the i-th generation (sometimes we will omit the generation index for the
third family, as we have done till now). Again here the normalization is arbitrary and can be
absorbed into the gauge coupling constant $g_F$. Since $\tr
Y_F=0$, all anomalies involving odd number of $U(1)_F$ cancel among
the three generations, as for instance those of the type
$U(1)_YU(1)_\chi U(1)_F$. Those involving $U(1)_F^2$ also cancel
because $Y_F^2$ is constant along all fields of a single generation.

Since the first two families have charges $Y_F=\pm 1/2$ while the
third generation has charge $Y_F=\pm 1$, the group $U(1)_F$ allows all
Yukawa couplings involving the first two generations, as well as
Yukawa couplings relating only third generation fermions. However the
mixing between the third and first or second generations is forbidden
and can only appear from non renormalizable operators as e.g.
$y^U_{3,i}\frac{\phi_F}{M_*} H_u Q_3 U^c_{i}$ ($i=1,2$) where $\phi_F$
is the field breaking $U(1)_F$.  In particular the most stringent
condition comes from the hierarchical structure of the fermion mass
matrix~\cite{Hall:1993ni} in the up sector which yields
$Y^U_{23}\equiv v_F\,y^U_{23} /M_*\simeq 10^{-1}$ and then
$v_F/M_*\gtrsim 10^{-2}$ for perturbative values of $y_{ij}^{U,D}$.
Such a constraint does not exist for $v_\chi$ as all Yukawa terms in
the Lagrangian are invariant under $U(1)_\chi$.

For simplicity, we will assume in this paper that both $U(1)_F$ and
$U(1)_\chi$ are spontaneously broken at most a few orders of magnitude
below $M_*$~\footnote{In this case it has been
  proved~\cite{Craig:2012yd} that the induced soft-breaking terms are
  indistinguishable from those of an unbroken gauge group.} by vector-like Higgses $\phi_F\,(\bar\phi_F)$ with $N_F Q_F$
charges $\pm\frac{1}{2}$ and $\phi_\chi\,(\bar\phi_\chi)$ with
$N_\chi Q_\chi$ charges $\pm q_\chi$.
These
breakings should be supersymmetric as in
Refs.~\cite{Craig:2012yd,Delgado:2011kr}, i.e. with a superpotential
as
\be 
\Delta W=\sum_{i=\chi,F} S_i(\bar\phi_i\phi_i-v_i^2)~,
\label{rotura}
\ee
where $S_i$ are singlets under the SM and $U(1)_\chi\otimes U(1)_F$
groups and $v_i$ are the breaking scales of the corresponding groups. 

The simplest model to transmit supersymmetry breaking from the hidden
to the observable sector consists in having messengers
$\Phi_\chi,\bar\Phi_\chi$ and $\Phi_F,\bar\Phi_F$ with charges $(\pm
Q_{\chi,\Phi_\chi},0)$ and $(0,\pm Q_{F,\Phi_F})$ under the group
$U(1)_\chi\otimes U(1)_F$, and with the superpotential
\be
\Delta W=\sum_{i=\chi,F}\left(\bar\Phi_i X\Phi_i+M_*\bar\Phi_i\Phi_i\right)
~ .
\ee
The messenger $\Phi_i$ ($i=\chi,F$) will transmit supersymmetry
breaking to the observable sector in which any generic sfermion
$\tilde f$ will receive at $\mathcal Q=M_*$ two-loop squared soft-mass terms as
\be
\Delta_i m^2_{\tilde f}\simeq
2\left(\frac{\alpha_i(t^*)}{4\pi}\right)^2 |Q_{i,\Phi_i}|^2~ |Q_{i,\tilde f}|^2~
\Lambda^2 ,\quad i=\chi,F
\label{scaM}
\ee
while the gaugino $\lambda_i$ receives at $\mathcal Q=M_*$ the one-loop Majorana mass
\be
M_i \simeq\, \frac{\alpha_i(t^*)}{4\pi} |Q_{i,\Phi_i}|^2
\Lambda ~,\quad i=\chi,F
\label{gauM}
\ee
which will in general translate into the inequality $M_\chi\neq M_F$
for $|Q_{\chi,\Phi_\chi}|\neq |Q_{F,\Phi_F}|$.

In conclusion the breaking of supersymmetry transmitted to the scalar
fields charged under the groups $U(1)_F$ and $U(1)_\chi$ can reproduce
the mass splitting between Higgses and squarks shown in Fig.~1. In
particular the boundary condition (\ref{UHu}) requires that the
coupling constants $\alpha_\chi$ and $\alpha_F$ at $\mathcal{Q}=M_*$
be related as
\be
\widetilde\alpha_F=\sqrt{\mathcal{R_\alpha}}\,\widetilde\alpha_\chi  ~
\label{relacion}
\ee
with $\mathcal R_\alpha=5/3$, where we have adopted the definitions
\be
\widetilde\alpha_\chi\equiv \frac{\alpha_\chi(t^*) \,Q_{\chi,\Phi_\chi}}{N_\chi},\quad
\widetilde\alpha_F\equiv \frac{\alpha_F(t^*) \,Q_{F,\Phi_F}}{N_F},
\label{definiciones}
\ee
so that from Eq.~(\ref{scaM}) any
sfermion $\tilde f$ acquires the mass $m^2_{\tilde f}(M_*)$ given by 
\be
\label{scalars}
m_{\tilde f}^2(M_*) = 
\frac{3}{8}\left(Q_{\chi,f}^2+\frac{5}{3} Q_{F,f}^2\right) m_{Q_3}^2(M_*)~
\ee
with
\bea
m^2_{Q_3}(M_*)= \frac{1}{3\pi^2}\left(\widetilde\alpha_\chi \Lambda\right)^2~.
\label{mQ0}\eea
Therefore, by using Eq.~(\ref{cero}) and neglecting the RGE evolution
of the masses whose running is tiny, one obtains the heavy-state mass
spectrum at $t=0$ in terms of the fundamental scale $m_Q\equiv
m_{Q_3}(0)$. This is presented in
\begin{table}[tb]
\begin{center}
\begin{tabular}{||c||c|c|c|c|c|c|c|c||}\hline
Field&$Q_3$&$E^c_3$&$Q_{1,2},\, U^c_{1,2},\, E^c_{1,2}$&$D^c_3,\, L_3$&$D^c_{1,2},\, L_{1,2}$&
$N^c_3$&$N^c_{1,2}$&$H^{0,\pm},\, A$\\
\hline
${\displaystyle \frac{m^2(0)}{m_Q^2}}$&1&2&17/16&8&113/16&20&305/16&3\\
\hline
\end{tabular}
\caption{\it Squared masses of the different heavy fields in units of $m_Q^2(0)$.}
\end{center}
\end{table} 
Tab.~2 that highlights a peculiar heavy mass pattern where the
lightest of the heavy states is the third generation squark doublet
$Q_3$.

Notice that the mechanism we have used to generate sfermion masses
does not involve the SM gauge group and thus does not give any
one-loop mass to the $SU(3)_c\otimes SU(2)_L\otimes U(1)_Y$
gauginos, nor it generates at one-loop any $A$-mixing
parameter, what is welcome for the strength of the electroweak phase
transition as we pointed out in the introduction. Of course there are
new gauginos $\lambda_\chi$ and $\lambda_F$ which get the Majorana
masses expressed in Eq.~(\ref{gauM}). We have not included their
effect in the RGE since we are assuming that $U(1)_\chi\otimes U(1)_F$
breaks only a few e-folds below $M_*$ and, since
$\alpha_{\chi}$ and $\alpha_{F}$ decrease quickly when decreasing the
scale, their impact on the previous results would be tiny.

Of course an alternative possibility is that $v_\chi$ be at the TeV
scale in which case an extra $Z$ gauge boson ($Z_\chi$) could be
present at low energy and might be detected at the LHC~\footnote{Due
  to the orthogonality of $U(1)_\chi$ with the SM gauge group,
  $Z_\chi$ has no kinetic mixing with the other gauge bosons. However,
  for this scenario with $v_\chi=\mathcal O$(TeV) we do not expect the
  spontaneous symmetry breaking of $U(1)_\chi$ to be supersymmetric as
  in (\ref{rotura}) but by some other mechanism, either at tree-level
  or by radiative corrections.}.  In that case we should add to the
$\beta$ functions in the RGE of Eqs.~(\ref{RGE}) the term
\be
\label{gaugi}
16\pi^2\Delta \beta_{m^2_{X}}=-8\left|Q_{\chi,\,X}\right|^2 g_\chi^2 M_\chi^2
\ee
where $X$ runs over all scalars of the theory. In any case the
influence of the term (\ref{gaugi}) when solving the RGE of
Eqs.~(\ref{RGE}) should be tiny since $U(1)_\chi$ is infrared free and
its impact on the RGE at low energy negligible. Nevertheless the
low-energy phenomenology of the model might be modified by
the presence of the $Z_\chi$ gauge boson.

\subsubsection{\sc The gaugino sector: gravity mediation}
\label{gaugino}

Besides the gauge mediators also gravity transmits supersymmetry
breaking to the visible sector. Indeed once supersymmetry is broken by
the vacuum expectation value $F$, minimization of the supergravity
potential leads to a mass $m_{3/2}$ for the gravitino. This mass is
fixed by the vanishing of the cosmological constant of the
supergravity potential which leads to the condition
\be
\left( \frac{F}{M_P}  \right)^2\simeq 3 m_{3/2}^2~,
\label{CC}
\ee
with $M_P=2.4\times 10^{18}$ GeV. Using Eqs.~(\ref{running}) and
(\ref{mQ0}) one can then express the gravitino mass as a function of the
fundamental scale $m_Q$ as
\be
m_{3/2}\simeq (m_{3/2})^0 \left( \frac{0.2 }{\widetilde\alpha_\chi} 
\right)\left( \frac{m_Q}{15\,\textrm{TeV}} \right)~,
\label{gravitino}
\ee
where $(m_{3/2})^0\simeq 170$ GeV. Thus gravity mediation introduces a
second fundamental scale $m_{3/2}$, much lower than $m_Q$, that will
be the natural scale of the gaugino sector. In fact gravity transmits to
gauginos (and scalars) masses of $\mathcal O(m_{3/2})$, which are
consistent with those of the LSS if the gravitino is at the electroweak
scale as we are assuming.  In addition gravity produces tri-linear
supersymmetry-breaking parameters $A$ generically of $\mathcal
O(m_{3/2})$ which implies $|A|^2 \ll m_Q^2$. This is in agreement with
the LSS because, as explained in the introduction, it restricts
the parameter space to the region where the EWBG requirement of a
strong electroweak phase transition is fulfilled.

The precise values of the soft-breaking terms induced by gravity
mediation depend on the particular model. For instance in a Polonyi
model~\cite{Polonyi} scalars receive a universal mass $m_0^2=m_{3/2}^2$,
gauginos receive masses $m_{1/2}=\mathcal O(m_{3/2})$, whose RGE
evolution to $t=0$ yield chargino and neutralino masses at the
gravitino scale and rises the gluino mass up to the TeV scale (because
of its strong renormalization) and $A=(3-\sqrt{3})m_{3/2}$. Of course
in other models these values could be changed and moreover one can
typically consider $m_0$, $m_{1/2}$ and $A$ as free parameters in supergravity models.

\subsubsection{\sc On the $\mu/B_\mu$-term generation} 
\label{muBmu}

We expect to generate the $\mu$ and $B_\mu$ parameters gravitationally
at the Planck scale as $\mu_1(M_P)\sim m_{3/2}$ and
$B_{\mu_1}(M_P)\sim m_{3/2}^2$ through Kahler-potential supergravity
terms~\cite{giudicemasiero}. For instance in a Polonyi-like model with
a scalar field $z$ spontaneously breaking local supersymmetry the
addition of the Kahler potential $\mathcal G=\sqrt{3}\lambda
(z^\dagger/M_P) H_uH_d$ generates $\mu$ and $B_\mu$ parameters given
by~\cite{giudicemasiero}:
\be
\mu_1(M_P)=\lambda (a-3)m_{3/2}~,\quad B_{\mu_1}(M_P)=\lambda
(2a-3)m_{3/2}^2~,\quad A(M_P)=a m_{3/2}~.
\label{PolKah}
\ee
Clearly the value of $\mu$ generated in this way (say $\mu=\mu_1$)
makes the masses of charginos and neutralinos to fit the ballpark
required by the LSS although the value of $B_{\mu_1}(M_P)$ would be
too small to trigger electroweak breaking.

On the other hand after integrating out the messengers of gauge
mediated supersymmetry breaking, two-loop diagrams from $U(1)_\chi$
gauge interactions give rise to a contribution to $B_{\mu_1}$ at the
scale $M_*$ as 
\be
B_{\mu_1}(M_*)\simeq \frac{\alpha_\chi(t^*)}{4\pi} |Q_{H_d}|^2\,M_\chi\,\mu_1~.
\ee
This value of $B_{\mu_1}(M_*)$ cannot
satisfy Eq.~(\ref{Bmu}) with $m_Q(0)\gtrsim 6$\,TeV and
$\tan\beta\lesssim 15$ as required by EDM and EWBG. This problem essentially
arises from the fact that $B_{\mu_1}$ is generated at two-loop and it
is proportional to $\mu_1$.

However this issue can be solved in gauge mediated theories where the
Peccei-Quinn global symmetry is explicitly broken by superpotential
interactions. In such theories either $B_\mu$ and $\mu$ receive 
one-loop contributions~\footnote{In fact when the messengers of gauge
  mediation are integrated out they contribute to the coefficients of
  the effective operators 
$$
\int d^4\theta \frac{ X^\dagger}{M_*}H_uH_d ~~, \quad
\int d^4\theta \frac{X^\dagger X}{M_*^2}H_uH_d ~~, \quad \cdots
\label{GM}
$$
which are expected to be of the same order of magnitude.}  $\mu_2$ and
$B_{\mu_2}$ that are related as~\cite{Dvali:1996cu}
\be
B_{\mu_2}(M_*)\simeq\mu_2\Lambda=\frac{\mu_2 F}{M_*} ~
\label{muBmu}
\ee
and then we can easily obtain~\footnote{Since an essential ingredient for EWBG is $\arg(\mu^* M_2)\neq 0$ we will be assuming, to simplify the forthcoming  numerical analysis, that $\mu$ is real while the $\arg(M_2)\neq 0$ should be provided by the gravity mediation mechanism.} 
$|B_{\mu_2}|\gg |B_{\mu_1}|$ with $|\mu_2|\lesssim|\mu_1|\sim m_{3/2}$, i.e.~$B_\mu\simeq B_{\mu_2}$ and $\mu\simeq \mu_1$. Clearly this relation is inconsistent with the
minimization condition~(\ref{Bmu}) if the soft-terms of the Higgs
sector are at the electroweak scale (the well-known $\mu/B_\mu$
problem) but it is consistent in the LSS where soft-terms in the Higgs
sector are in the multi-TeV region as we will see now.
\begin{figure}[htb]
\begin{center}
\includegraphics[width=0.45\textwidth]{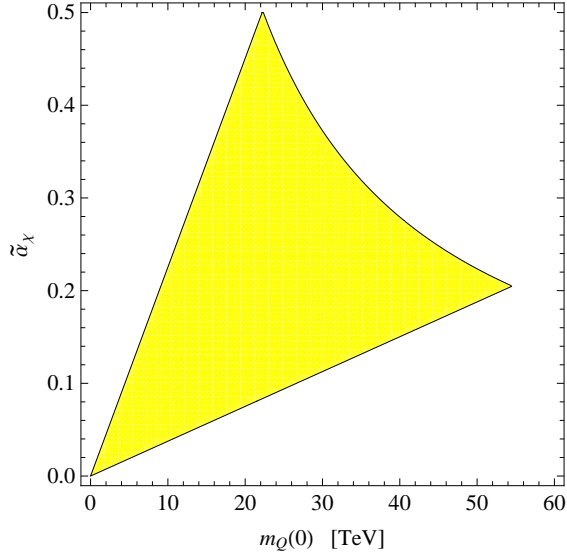}
\end{center}
\caption{\it Allowed region in the plane $[\widetilde\alpha_\chi,m_Q(0)]$. The upper (lower) straight-line border of the allowed cone corresponds to the bound $m_{3/2}=100\ (600) $ GeV and the border parabola to the bound (\ref{condiciones}). 
}
\label{plot2}
\end{figure}
In fact from Eq.~(\ref{muBmu}) and using the results from the previous
sections one obtains
\be
|B_{\mu_2}|\simeq(B_{\mu_2})^0 \left( \frac{0.2 }{\widetilde\alpha_\chi} \right)\left( \frac{m_Q(0)}{15\,\textrm{TeV}} \right)
\left( \frac{|\mu_2|}{100\,\textrm{GeV}} \right)
\label{B}
\ee
where $(B_{\mu_2})^0\simeq(7.7\, \textrm{TeV})^2$ while the minimization
condition (\ref{Bmu}) provides a relation on the different parameters
as
\be
|\mu_2|\simeq (\mu_2)^0\left(\frac{15}{\tan\beta}\right) \left( \frac{\widetilde\alpha_\chi}{0.2} \right)
\left( \frac{m_Q(0)}{15\,\textrm{TeV}} \right)
\label{tan}
\ee
where $|(\mu_2)^0|\simeq 80$ GeV. Since to have a successful EWBG scenario
we require light charginos and neutralinos we will impose on
(\ref{tan}) the constraints
\be
\tan\beta<15~,\quad |\mu_2|<300\ \textrm{GeV}\ .
\label{condiciones}
\ee
By imposing the condition $100\mbox{ GeV}<m_{3/2}<600\mbox{ GeV}$ in
(\ref{gravitino}) and (\ref{tan}) we can constrain
$\widetilde\alpha_\chi$ and $m_Q(0)$ as exhibited in Fig.~2.  Notice
that $\widetilde\alpha_\chi\sim \mathcal O(1)$ in the upper
corner of the allowed region although it runs down very quickly at
lower scales $M<M_*$. On the other hand we can see that for values
$m_Q\simeq 10- 20$ TeV there is a wide region where the values of the
gauge coupling $\widetilde\alpha_\chi$ are in better agreement with
perturbativity. Finally small values of $m_Q(0)$ and
$\widetilde\alpha_\chi$ correspond to small values of $\mu_2$ and
therefore for such values the $\mu$ term has to be mostly due to
gravity mediation for neutralinos and charginos to be in the ballpark
of the LSS.

\subsubsection{\sc The Higgs mass in the minimal model}
\label{considerations}

The low energy effective theory below the scale $m_Q$ of the model we
have just presented contains the SM fields, gluinos, light charginos
and neutralinos and the right-handed stop lighter than the top. The
SM-Higgs mass is fixed by its effective quartic coupling and by the
radiative corrections provided by the light
fields~\cite{Carena:1995wu,Carena:2008rt}. In turn the effective
quartic coupling is fixed at the scale $m_Q$ by the supersymmetric
quartic coupling and by threshold effects depending on the top mixing
$A_t$ of the underlying supersymmetric theory. To satisfy the
requirements of EWBG the mixing $A_t/m_Q(0)$ should be small and
indeed it is negligible in this minimal model since: i) The gauge
mediation of supersymmetry breaking can only provide mixing parameters
that are suppressed with respect to the value of the scalar masses;
ii) Gravity mediation provides mixing parameters of the order of the
gravitino mass, which are then much smaller than the heavy scalar
masses. In fact we have consistently neglected the influence of the
mixing $A_t$ when running the RGE. In conclusion the minimal model we
have just presented proves that a fundamental theory reproducing the
LSS exists. However the example we have provided covers only part of
the possible parameter space of the LSS namely the region with
negligible $A_t$~\footnote{One could of course envisage models, with
  non-generic Kahler potential and superpotential for scalar fields
  spontaneously breaking local supersymmetry, yielding values $A\gg
  m_{3/2}$ in the low-energy effective
  supersymmetric theory.  For instance this happens in the case of
  Eq.~(\ref{PolKah}) for $m_{3/2}\sim 100$ GeV, $a\sim 10$ and 
  $\lambda\sim 0.1$ to keep $\mu$ at the electroweak scale. The same
  also occurs for $m_{3/2}\sim$TeV, $a\sim 3$ and $\lambda\sim
  0.1$. On the other hand such cases eventually reach $A_t\approx
  m_Q(0)/2$ only for small values of $m_Q(0)$ and they would require
  general supergravity couplings, whose study is outside the scope of
  the present paper.}. Under these circumstances one expects to find
an upper bound on the SM-Higgs mass that is stronger than in the
general LSS case (where successful EWBG predicts $m_h\lesssim 127$
GeV~\cite{Carena:2008rt}). In fact using the approach of
Ref.~\cite{Carena:2008rt} the proposed minimal model predicts
$m_h\lesssim 115.5$ GeV for the allowed region of
Fig.~\ref{plot2}. Then the predicted Higgs mass barely overcomes the
recent LHC constraints~\footnote{Of course this
  bound is slightly relaxed if EWBG is not enforced (see
  footnote~\ref{footnoteLSS}).}.

\subsection{\sc A non-minimal UV completion}
\label{nominimal}
If the Higgs mass turns out to be around 125 GeV, as the excesses
found by the ATLAS and CMS Collaborations seem to indicate, then we
need to modify the above UV completion without altering its main EWBG
capabilities nor the effective theory which constitutes the essence of
the LSS. A simple way of achieving that is by introducing a pair of
vector-like extra multiplets, charged under the gauge group
$U(1)_\chi$, which enhance the effective quartic coupling of the
SM-like Higgs.

Hence, in addition to the MSSM superfield content, we consider two
$SU(2)$ triplets $T_{d,u}$ with superpotential
\be 
\Delta W=\chi_u H_u\cdot T_u H_u+\chi_d H_d\cdot T_d H_d +
\mu_T T_d T_u~.
\label{superT}
\ee 
$T_{d,u}$ have charges $Y=\pm 1$ and $Q_\chi=\pm 4/N_\chi$ under the
$U(1)_Y$ and $U(1)_\chi$ gauge groups while they are neutral under
$U(1)_F$.  The couplings $\chi_{u,d}$ modify the beta-functions
$\beta_{h_t},\beta_{h_b}$ and $\beta_{g_3}$ \cite{Espinosa:1991gr} as
well as the RGE (\ref{RGE}) as
\begin{eqnarray}
16\pi^2\Delta\beta_{m_{H_u}^2}&=&12\,\chi_u^2 Y_U~,\quad 16\pi^2\Delta\beta_{m_{H_d}^2}=12\,\chi_d^2 Y_D~,\nn\\
16\pi^2\beta_{m_{T_u}^2}&=&4 \chi_u^2 Y_U~,\quad 16\pi^2\beta_{m_{T_d}^2}=4 \chi_d^2 Y_D~,
\label{RGET}
\end{eqnarray}
where
\be
\label{Ys}
Y_U= m_{T_u}^2+2 m_{H_u}^2~,\quad Y_D= m_{T_d}^2+2 m_{H_d}^2~.
\ee
To some extent, and depending
on the actual values of the couplings $\chi_{u,d}$, the extra terms in the RGE (\ref{RGET}) distort the analysis of Sec.~\ref{renormalization}. In particular the RGE
invariants in Eqs.~(\ref{I}) get modified as
\begin{eqnarray}
\Delta I_Q(t)&=& 3 m_{T_u}^2+3 m_{T_d}^2~,\nn\\
\Delta I_U(t)&=& 3 m_{T_u}^2~,\\
\Delta I_D(t)&=& 3 m_{T_d}^2~.\nn
\label{IT}
\end{eqnarray}
However the analysis of Sec.~\ref{renormalization} can be done
straightforwardly in the presence of the new couplings $\chi_{u,d}$
and the scalars of $T_{u,d}$, which acquire a soft-mass
$m_{T_{u,d}}^2(t^*) = 4\, m_{H_u}^2(t^*)$ by $U(1)_\chi$ gauge
mediation.  The fermions in $T_{u,d}$ will mix with charginos and
neutralinos after electroweak symmetry breaking while the mass $\mu_T$
in the superpotential can be generated by some $D$-term effective
operators~\cite{giudicemasiero}. We do not expect this mixing should
alter the generation of $CP$-violating currents required by EWBG.

As it has been already explained in Sec.~\ref{minimal} the gauge and
gravity mediation mechanism generically generate a tiny value of
$A_t/m_Q$ due to the hierarchy $m_{3/2}\ll m_Q$. However this feature
does not imply a stringent upper bound on $m_h$ unlike in the minimal
model. Indeed the $F$-terms of the superpotential (\ref{superT})
provide an additional contribution to the quartic coupling of the
SM-like Higgs and the tree-level Higgs mass gets enhanced by the
amount $\Delta m_h^2\simeq 4 \chi_u^2 \, v^2$ (for
$\tan\beta\gg1$)~\cite{Espinosa:1991gr}. Hence if the Higgs mass
turned out to be just beyond the capabilities of the minimal model, a
small value of $\chi_u$ would be enough to reproduce the Higgs mass
value, and the results of Figs.~1 and 2 would still be valid. On the
other hand for a larger Higgs mass $\chi_u$ should be
sizeable~\footnote{In general one must worry about the Landau pole
  that this extra coupling can
  generate~\cite{Espinosa:1991gr}. However the required values of
  $\chi_u$ ($\chi_u\lesssim 0.15$) do not create any problem.} and the
contributions (\ref{RGET}) would be important.

As an illustrative case we perform the numerical analysis of the model
for $\chi_u=0.15$ such that the prediction of the Higgs mass, for the
values of $m_Q$ we will obtain, will cover the upper part of the
experimentally-allowed region $115.5\,\textrm{GeV}\lesssim m_h\lesssim
127\,\rm$GeV \cite{lhc}. Moreover we set $\chi_d=0$ because it plays a
minor role in the determination of the Higgs mass in the large
$\tan\beta$ regime.  

The solution of the RGE gives now $t^*\simeq 24.1 $,
$M_*\simeq2.6\times 10^{12}\mbox{ GeV}$, while the coefficients
$\mathcal R_m$ and $\mathcal R_\alpha$ defined in Eqs.~(\ref{UHu}) and
(\ref{relacion}), respectively, take now the values $\mathcal
R_m\simeq 1/3$ and $\mathcal R_\alpha\simeq 2/7$ and the heavy field
spectrum given in Tab.~2 for the case of the minimal UV completion is
now deformed to that in Tab.~3.
%
%
%

%
\begin{table}[htb]
\begin{center}
\begin{tabular}{||c||c|c|c|c|c|c|c|c|c|c||}\hline
Field&$Q_3$&$E^c_3$&$Q_{1,2},\, U^c_{1,2},\, E^c_{1,2}$&$D^c_3,\, L_3$&$D^c_{1,2},\, L_{1,2}$&
$N^c_3$&$N^c_{1,2}$&$H^{0,\pm},\, A$& $T_u$& $T_d$\\
\hline
${\displaystyle \frac{m(0)}{m_Q}}$&1&1.5&1.3&4.0&3.9&6.6&6.5&2.6&5.1&5.2\\
\hline
\end{tabular}
\caption{\it Masses of the different heavy fields in units of $m_Q(0)$ for the case of non-minimal UV completion with $\chi_u=0.15$ and $\chi_d=0$.}
\end{center}
\end{table} 
\begin{figure}[h!]
\begin{center}
\includegraphics[width=0.41\textwidth]{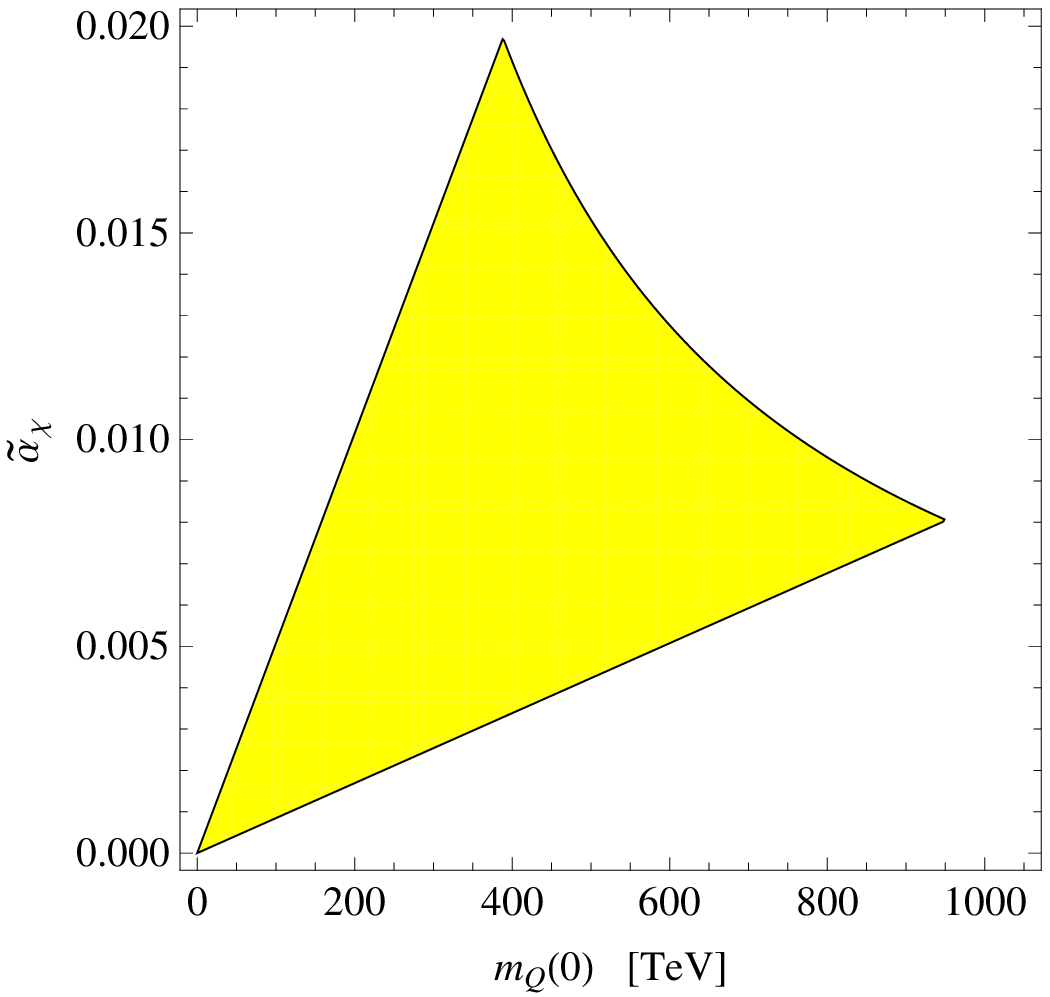}\qquad
\includegraphics[width=0.5\textwidth]{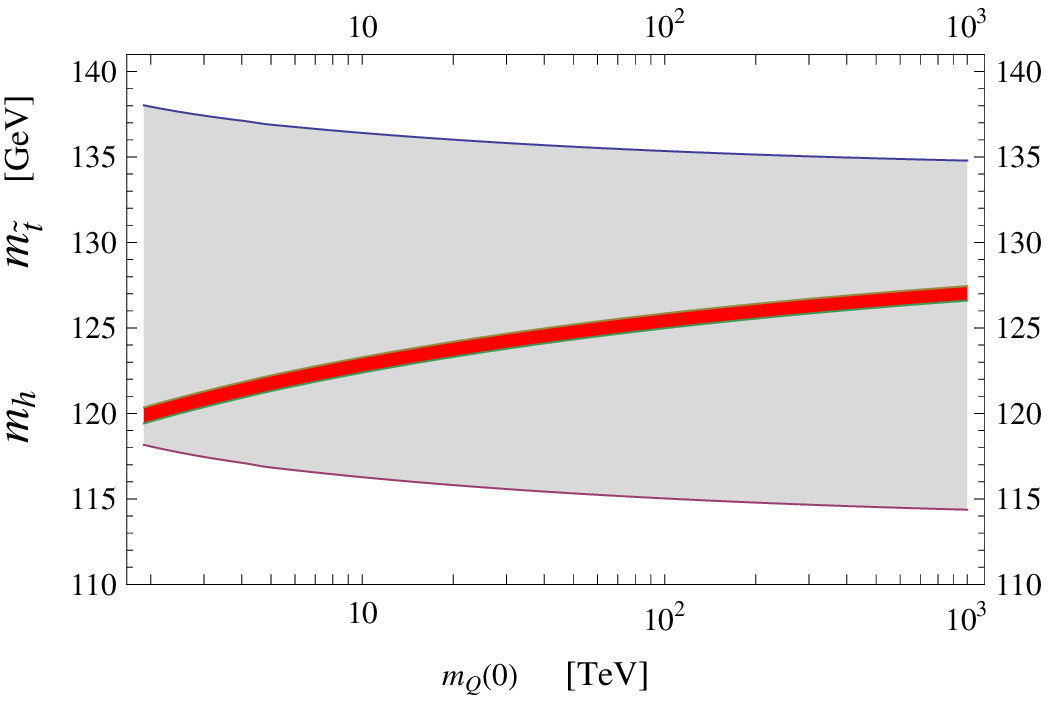}\\
\end{center}
\caption{\it Left panel: Allowed region in the plane
  $[\widetilde\alpha_\chi,m_Q(0)]$ for $\chi_u=0.15$. The upper
  (lower) straight-line border of the allowed cone corresponds to the
  bound $m_{3/2}=100\ (600)$ GeV and the border parabola to the bound
  (\ref{condiciones}). Right panel: the Higgs mass $m_h$ (red band) and
  the lightest stop mass $m_{\widetilde t}$ (gray band) as a function
  of $m_Q$ in the non-minimal UV completion model with $\chi_u=0.15$,
  $\chi_d=0$, $\tan\beta=10$ and $(100\,{\rm GeV})^2\!>-m_U^2\!>(70\,{\rm
    GeV})^2$.}
\label{triplete}
\end{figure}

The coefficients in Eqs.~(\ref{gravitino}) and (\ref{tan}) become $(m_{3/2})^0\simeq
0.38$ GeV and $(\mu_2)^0\simeq 117$ GeV. As in the previous analysis
we can restrict the gravitino mass and $\tan\beta$ to the intervals
$100\,\textrm{GeV}<m_{3/2}<600\,\textrm{GeV}$ and $\tan\beta<15$. For
such restriction and imposing $|\mu_2|<300\,$GeV the allowed region
for the parameters $\widetilde\alpha_\chi$ and $m_Q(0)$ is the one shown in
Fig.~\ref{triplete} (left panel).

Notice that the allowed values of $\widetilde\alpha_\chi$ are now well
inside the perturbative region.  For $\chi_u=0.15$ the model is
compatible with any SM-like Higgs mass up to 127 GeV. This is
exhibited in Fig.~\ref{triplete} (right panel) where the masses are
calculated in the approach of Ref.~\cite{Carena:2008rt} with the
addition of the new coupling $\chi_u$ in the matching conditions
of the effective quartic coupling $\lambda$ of the SM-like Higgs:
\be
\lambda(m_Q) = \frac{g^2(m_Q)+g^{\prime
      2}(m_Q)+16 \chi_u^2(m_Q)} {4} \cos^22\beta \left(1-{1\over2}\Delta
      Z_\lambda\right)+\Delta\lambda~, \label{lambda}\\ 
\ee
where $\Delta Z_\lambda$ and $\Delta\lambda$ are respectively the
finite wave-function and proper-vertex threshold corrections due to
the heavy scalar decoupling~\cite{Carena:2008rt}.
Of course for $\chi_u>0.15$ smaller values of $m_Q$ are required to
get the same Higgs masses while smaller values of $m_h$ are also
allowed with couplings $\chi_u<0.15$.

\section{\sc Conclusion and Outlook}
\label{conclusions}
In this paper we have built up a high energy model of supersymmetry
breaking giving rise at the electroweak scale to the so-called Light
Stop Scenario of the MSSM where Electroweak Baryogenesis can reproduce
the observed baryon asymmetry of the Universe. The model contains as
light fields the Standard Model fields plus the right-handed stop
$\tilde t$ (to trigger a strong enough electroweak phase transition)
and charginos and neutralinos (to generate the $CP$ violating currents
responsible of the baryon asymmetry). The main feature of the LSS is
that the right handed stop is lighter than the top quark $m_{\tilde
  t}<m_t$. Such a light stop being an $SU2)_L$ singlet does not change
significantly the Standard Model precision observables and moreover it
is not excluded by experimental searches at Tevatron and LHC provided
that the lightest neutralino can be produced on-shell in the decay
$\tilde t\to c\, \chi^0$ yielding a lower bound $m_{\tilde t}\gtrsim
95$ GeV~\cite{stopCDF}.

A first observation which could be made is that in order to trigger
electroweak breaking and small values of the right-handed stop mass
one requires very heavy squark doublet $Q$ at low energy. By evolving
the RGE this implies heavy values for all scalar masses at high
scales. This, along with the existence of light gauginos and
Higgsinos, strongly suggests a mechanism of supersymmetry-breaking
transmission mediated by non-Standard Model gauge interactions under
which the MSSM chiral superfields are charged. Dictated by the RGE
evolution we have identified such a gauge group as the product of an
extra $U(1)_F$, a subgroup of $SU(3)_F$ acting on the three families,
and the well known $U(1)_\chi$ which for instance arises in the
breaking of $SO(10)\to SU(5)\otimes U(1)\chi$ and does not necessitate
extra matter, apart from right-handed neutrinos, to cancel anomalies.
From the results of RGE evolution this supersymmetry breaking should
be mediated by messengers with supersymmetric mass of order
$M_*\lesssim10^{15}$~GeV. This, along with the mass spectrum of the
scalars, can fix the gravitino mass in the $100-600\,\rm GeV$
range. By Planck-scale suppressed interactions this supersymmetry
breaking is then transmitted to the observable sector, and in
particular to gauginos and Higgsinos, which then acquire
supersymmetry-breaking masses at the electroweak scale as required by
the EWBG mechanism. The texture of Yukawa couplings requires that
$U(1)_F$ break at most a few orders of magnitude below $M_*$. However
$U(1)_\chi$ can break down at the TeV range and the gauge boson
$Z_\chi$, a remnant of the UV completion of the LSS, could be detected
at LHC. Thus the existence of an extra (heavy or superheavy) gauge
boson $Z_\chi$ is predicted by the UV completion of the LSS. However
in this paper we have only considered the case of superheavy gauge
boson $Z_\chi$, which does not modify electroweak precision
observables and is undetectable at LHC.

Consistency with electroweak breaking conditions then yields an upper
bound on the mass of left-handed stops as $m_Q\lesssim
60\,\textrm{TeV}$ which is in agreement (though in the low range) with
the values usually assumed in EWBG
calculations~\cite{Carena:2008rt}. In addition the model can generate
only small stop mixing $A_t/m_Q$, as preferred by the EWBG requirement
of a strong first-order phase transition, and then reproduce only the
part of the LSS parameter space where the SM-like Higgs mass barely
overcomes the experimental bound. Thus further investigations to
reproduce the whole LSS parameter space (namely the EWBG region with
$A_t\lesssim m_Q/2$ and $m_Q\gtrsim 10$\,TeV) are still needed. In
this sense considering general supergravity couplings in Planck
mediated supersymmetry breaking seems promising.

Finally if the recent excess found by the ATLAS and CMS
Collaborations at values of the Higgs mass around 120--125 GeV gets
confirmed one can extend the above supersymmetry-breaking model by
introducing some extra matter which modifies the Higgs quartic
coupling without altering the scalar content of the LSS. We have
presented a simple model involving a pair of vector-like $Y=\pm 1$
triplets for which the Higgs mass value can easily cover the LHC
allowed region $115.5\,\textrm{GeV}\lesssim m_H\lesssim
127\,\textrm{GeV}$. This consists in a different ultraviolet
completion but in an effective low energy theory with the same
scalar content of the LSS and a richer fermionic sector coming from
the mixing of the new triplets with gauginos and Higgsinos. The
prospects of such an effective theory for EWBG are worth
investigating.

\section*{\sc Acknowledgments}
GN thanks the University of Notre Dame and the Argonne National
Laboratory for their kind hospitality during the first stages of this
work.  AD was partly supported by the National Science Foundation
under grant PHY-0905383-ARRA. MQ was supported in part by the Spanish
Consolider-Ingenio 2010 Programme CPAN (CSD2007-00042) and by
CICYT-FEDER-FPA2008-01430 and FPA2011-25948.


\begin{thebibliography}{99}



\bibitem{Sakharov:1967dj}
  A.~D.~Sakharov,
  Pisma Zh.\ Eksp.\ Teor.\ Fiz.\  {\bf 5}, 32 (1967)
  [JETP Lett.\  {\bf 5}, 24 (1967)]
  [Sov.\ Phys.\ Usp.\  {\bf 34}, 392 (1991)]
  [Usp.\ Fiz.\ Nauk {\bf 161}, 61 (1991)].



\bibitem{Kuzmin:1985mm}
  V.~A.~Kuzmin, V.~A.~Rubakov and M.~E.~Shaposhnikov,
  Phys.\ Lett.\ B {\bf 155}, 36 (1985).

\bibitem{Cohen:1993nk}
  See, e.g.: A.~G.~Cohen, D.~B.~Kaplan and A.~E.~Nelson,
  Ann.\ Rev.\ Nucl.\ Part.\ Sci.\ {\bf 43}, 27 (1993)
  [hep-ph/9302210].

\bibitem{AndH} 
G.~W.~Anderson and L.~J.~Hall,
  Phys.\ Rev.\  D {\bf 45}, 2685 (1992).

%
\bibitem{improvement}
M.~E.~Carrington,
  Phys.\ Rev.\  D {\bf 45}, 2933 (1992);
  M.~Dine, R.~G.~Leigh, P.~Huet, A.~D.~Linde and D.~A.~Linde,
  Phys.\ Lett.\  B {\bf 283}, 319 (1992)
  [arXiv:hep-ph/9203201];
  Phys.\ Rev.\  D {\bf 46}, 550 (1992)
  [arXiv:hep-ph/9203203];
  J.~R.~Espinosa, M.~Quiros and F.~Zwirner,
  Phys.\ Lett.\  B {\bf 314}, 206 (1993)
  [arXiv:hep-ph/9212248];
  W.~Buchmuller, Z.~Fodor, T.~Helbig and D.~Walliser,
  Annals Phys.\  {\bf 234}, 260 (1994)
  [arXiv:hep-ph/9303251].

%
\bibitem{twoloop} 
P.~Arnold and O.~Espinosa,
  Phys.\ Rev.\  D {\bf 47}, 3546 (1993)
  [Erratum-ibid.\  D {\bf 50}, 6662 (1994)]
  [arXiv:hep-ph/9212235].
%
\bibitem{nonpert}
K.~Kajantie, K.~Rummukainen and M.~E.~Shaposhnikov,
  Nucl.\ Phys.\  B {\bf 407}, 356 (1993)
  [arXiv:hep-ph/9305345];
  Z.~Fodor, J.~Hein, K.~Jansen, A.~Jaster and I.~Montvay,
  Nucl.\ Phys.\  B {\bf 439}, 147 (1995)
  [arXiv:hep-lat/9409017];
  K.~Kajantie, M.~Laine, K.~Rummukainen and M.~E.~Shaposhnikov,
  Nucl.\ Phys.\  B {\bf 466}, 189 (1996)
  [arXiv:hep-lat/9510020];
  K.~Jansen,
  Nucl.\ Phys.\ Proc.\ Suppl.\  {\bf 47}, 196 (1996)
  [arXiv:hep-lat/9509018].
For an alternative approach, see:
 B.~Bergerhoff and C.~Wetterich,
  Nucl.\ Phys.\  B {\bf 440}, 171 (1995)
  [arXiv:hep-ph/9409295] and references therein.

%
\bibitem{CPSM}
 G.~R.~Farrar and M.~E.~Shaposhnikov,
  Phys.\ Rev.\ Lett.\  {\bf 70}, 2833 (1993)
  [Erratum-ibid.\  {\bf 71}, 210 (1993)]
  [arXiv:hep-ph/9305274];
  M.~B.~Gavela, P.~Hernandez, J.~Orloff and O.~Pene,
  Mod.\ Phys.\ Lett.\  A {\bf 9}, 795 (1994)
  [arXiv:hep-ph/9312215];
  M.~B.~Gavela, P.~Hernandez, J.~Orloff, O.~Pene and C.~Quimbay,
  Nucl.\ Phys.\  B {\bf 430}, 382 (1994)
  [arXiv:hep-ph/9406289];
  P.~Huet and E.~Sather,
  Phys.\ Rev.\  D {\bf 51}, 379 (1995)
  [arXiv:hep-ph/9404302].

%
%
\bibitem{early} 
 G.~F.~Giudice,
  Phys.\ Rev.\  D {\bf 45}, 3177 (1992);
  K.~S.~Myint,
  Nucl.\ Phys.\  A {\bf 547}, 227C (1992);
 J.~R.~Espinosa, M.~Quiros and F.~Zwirner,
  Phys.\ Lett.\  B {\bf 307}, 106 (1993)
  [arXiv:hep-ph/9303317];
A.~Brignole, J.~R.~Espinosa, M.~Quiros and F.~Zwirner,
  Phys.\ Lett.\  B {\bf 324}, 181 (1994)
  [arXiv:hep-ph/9312296].

\bibitem{CQW} 
M.~S.~Carena, M.~Quiros and C.~E.~M.~Wagner,
  Phys.\ Lett.\  B {\bf 380}, 81 (1996)
  [arXiv:hep-ph/9603420].

%
\bibitem{Delepine} 
D.~Delepine, J.~M.~Gerard, R.~Gonzalez Felipe and J.~Weyers,
  Phys.\ Lett.\  B {\bf 386}, 183 (1996)
  [arXiv:hep-ph/9604440].


%
\bibitem{CK} 
 J.~M.~Cline and K.~Kainulainen,
  Nucl.\ Phys.\  B {\bf 482}, 73 (1996)
  [arXiv:hep-ph/9605235];
  Nucl.\ Phys.\  B {\bf 510}, 88 (1998)
  [arXiv:hep-ph/9705201];
 M.~Laine and K.~Rummukainen,
  Nucl.\ Phys.\  B {\bf 535}, 423 (1998)
  [arXiv:hep-lat/9804019];
  Phys.\ Rev.\ Lett.\  {\bf 80}, 5259 (1998)
  [arXiv:hep-ph/9804255].
%
\bibitem{FL} 
 M.~Laine,
  Nucl.\ Phys.\  B {\bf 481}, 43 (1996)
  [Erratum-ibid.\  B {\bf 548}, 637 (1999)]
  [arXiv:hep-ph/9605283];
M.~Losada,
  Phys.\ Rev.\  D {\bf 56}, 2893 (1997)
  [arXiv:hep-ph/9605266]; 
  preprint arXiv:hep-ph/9612337;
G.~R.~Farrar and M.~Losada,
  Phys.\ Lett.\  B {\bf 406}, 60 (1997)
  [arXiv:hep-ph/9612346]. 
%
\bibitem{JoseR}
J.~R.~Espinosa,
  Nucl.\ Phys.\  B {\bf 475}, 273 (1996)
  [arXiv:hep-ph/9604320].
%
\bibitem{JRB} 
 B.~de Carlos and J.~R.~Espinosa,
  Nucl.\ Phys.\  B {\bf 503}, 24 (1997)
  [arXiv:hep-ph/9703212].
%
\bibitem{Carena:1997gx}
M.~S.~Carena, M.~Quiros, A.~Riotto, I.~Vilja and C.~E.~M.~Wagner,
  Nucl.\ Phys.\  B {\bf 503}, 387 (1997)
  [arXiv:hep-ph/9702409].

%
\bibitem{Carena:1997ki}
  M.~S.~Carena, M.~Quiros and C.~E.~M.~Wagner,
  Nucl.\ Phys.\  B {\bf 524}, 3 (1998)
  [arXiv:hep-ph/9710401].
%
\bibitem{CJK}
 J.~M.~Cline, M.~Joyce and K.~Kainulainen,
  Phys.\ Lett.\  B {\bf 417}, 79 (1998)
  [Erratum-ibid.\  B {\bf 448}, 321 (1999)]
  [arXiv:hep-ph/9708393].
%
\bibitem{Iiro2} 
T.~Multamaki and I.~Vilja,
  Phys.\ Lett.\  B {\bf 411}, 301 (1997)
  [arXiv:hep-ph/9705469].
%
\bibitem{Toni2} 
A.~Riotto,
  Int.\ J.\ Mod.\ Phys.\  D {\bf 7}, 815 (1998)
  [arXiv:hep-ph/9709286].
%
\bibitem{Worah} 
 M.~P.~Worah,
  Phys.\ Rev.\  D {\bf 56}, 2010 (1997)
  [arXiv:hep-ph/9702423].
%
\bibitem{Schmidt}
  D.~Bodeker, P.~John, M.~Laine and M.~G.~Schmidt,
  Nucl.\ Phys.\  B {\bf 497}, 387 (1997)
  [arXiv:hep-ph/9612364].
%
\bibitem{Cline:2000kb}
  J.~M.~Cline and K.~Kainulainen,
  Phys.\ Rev.\ Lett.\  {\bf 85}, 5519 (2000)
  [arXiv:hep-ph/0002272];
 J.~M.~Cline, M.~Joyce and K.~Kainulainen,
  JHEP {\bf 0007}, 018 (2000)
  [arXiv:hep-ph/0006119].

\bibitem{Carena:2000id}
  M.~S.~Carena, J.~M.~Moreno, M.~Quiros, M.~Seco and C.~E.~M.~Wagner,
  Nucl.\ Phys.\  B {\bf 599}, 158 (2001)
  [arXiv:hep-ph/0011055];
 M.~S.~Carena, M.~Quiros, M.~Seco and C.~E.~M.~Wagner,
  Nucl.\ Phys.\  B {\bf 650}, 24 (2003)
  [arXiv:hep-ph/0208043].

\bibitem{Konstandin:2005cd}
  T.~Konstandin, T.~Prokopec, M.~G.~Schmidt and M.~Seco,
  Nucl.\ Phys.\ B {\bf 738}, 1 (2006)
  [arXiv:hep-ph/0505103].

\bibitem{Cirigliano:2006dg}
 C.~Lee, V.~Cirigliano and M.~J.~Ramsey-Musolf,
  Phys.\ Rev.\ D {\bf 71}, 075010 (2005)
  [hep-ph/0412354];
  V.~Cirigliano, S.~Profumo and M.~J.~Ramsey-Musolf,
  JHEP {\bf 0607}, 002 (2006)
  [arXiv:hep-ph/0603246];
  S.~Y.~Ayazi,
  eConf C {\bf 0605151}, 0004 (2006)
  [hep-ph/0611056];
  V.~Cirigliano, Y.~Li, S.~Profumo and M.~J.~Ramsey-Musolf,
  JHEP {\bf 1001}, 002 (2010)
  [arXiv:0910.4589 [hep-ph]].
 

\bibitem{Carena:2008rt}
  M.~Carena, G.~Nardini, M.~Quiros and C.~E.~M.~Wagner,
  JHEP {\bf 0810}, 062 (2008)
  [arXiv:0806.4297 [hep-ph]];
  Nucl.\ Phys.\  B {\bf 812}, 243 (2009)
  [arXiv:0809.3760 [hep-ph]].

\bibitem{Martin:1993zk} 
  S.~P.~Martin and M.~T.~Vaughn,
  Phys.\ Rev.\ D {\bf 50}, 2282 (1994)
  [Erratum-ibid.\ D {\bf 78}, 039903 (2008)]
  [hep-ph/9311340].
%

\bibitem{lhc}
{ATLAS Collaboration}, {Combination of Higgs Boson Searches with up to
  4.9~fb$^{-1}$ of pp Collisions Data Taken at a center-of-mass energy of 7~TeV
  with the ATLAS Experiment at the LHC}, 
  \href{http://cdsweb.cern.ch/record/1406358}{ATLAS-CONF-2011-163};
{CMS Collaboration}, {Combination of SM Higgs Searches}, 
  \href{http://cdsweb.cern.ch/record/1406347/}{CMS-PAS-HIG-11-032}.



\bibitem{gmsb}
   G.~F.~Giudice, R.~Rattazzi,
  Phys.\ Rept.\  {\bf 322}, 419-499 (1999)
  [hep-ph/9801271].

\bibitem{Langacker:2008yv} 
  P.~Langacker,
  Rev.\ Mod.\ Phys.\  {\bf 81}, 1199 (2009)
  [arXiv:0801.1345 [hep-ph]].

\bibitem{paco} 
  F.~del Aguila, M.~Quiros and F.~Zwirner,
  Nucl.\ Phys.\ B {\bf 287}, 419 (1987).

\bibitem{Hall:1993ni}
  L.~J.~Hall, A.~Rasin,
  Phys.\ Lett.\  {\bf B315}, 164-169 (1993)
  [hep-ph/9303303].



\bibitem{Craig:2012yd} 
   E.~Gorbatov and M.~Sudano,
  JHEP {\bf 0810}, 066 (2008)
  [arXiv:0802.0555 [hep-ph]];
    N.~Craig, M.~McCullough and J.~Thaler,
  arXiv:1201.2179 [hep-ph].

\bibitem{Delgado:2011kr} 
  A.~Delgado and M.~Quiros,
  arXiv:1111.0528 [hep-ph].

\bibitem{Polonyi}
 H.~P.~Nilles,
  Phys.\ Rept.\  {\bf 110}, 1-162 (1984).
 
\bibitem{giudicemasiero}
 G.~F.~Giudice, A.~Masiero,
  Phys.\ Lett.\  {\bf B206}, 480-484 (1988).

\bibitem{Carena:1995wu}
  M.~S.~Carena, M.~Quiros, C.~E.~M.~Wagner,
  Nucl.\ Phys.\  {\bf B461}, 407-436 (1996)
  [hep-ph/9508343].


\bibitem{Dvali:1996cu} 
  G.~R.~Dvali, G.~F.~Giudice and A.~Pomarol,
  Nucl.\ Phys.\ B {\bf 478}, 31 (1996)
  [hep-ph/9603238].

\bibitem{Espinosa:1991gr} 
  J.~R.~Espinosa and M.~Quiros,
  Phys.\ Lett.\ B {\bf 279}, 92 (1992);
  Phys.\ Rev.\ Lett.\  {\bf 81}, 516 (1998)
  [hep-ph/9804235].

\bibitem{stopCDF}
  P.~Calfayan [CDF and D0 Collaboration],
  AIP Conf.\ Proc.\  {\bf 1078} (2009) 262.


  \end{thebibliography}
\end{document}